\documentclass
[pra,letterpaper,10pt,showpacs,superscriptaddress,twocolumn]{revtex4}%
\usepackage{amsfonts}
\usepackage{amsmath}
\usepackage{amssymb}
\usepackage{graphicx}%
\usepackage{color}
\providecommand{\U}[1]{\protect\rule{.1in}{.1in}}

\begin{document}
\title{Ultralong-range polyatomic Rydberg molecules formed by a polar perturber}
\author{Seth T. Rittenhouse}
\affiliation{ITAMP, Harvard-Smithsonian Center for Astrophysics, Cambridge, MA 02138}
\author{M. Mayle}
\altaffiliation[Present address: ]{JILA, University of Colorado and National Institute of Standards and Technology, Boulder, Colorado 80309-0440, USA.}
\affiliation{Zentrum f\"{u}r Optische Quantentechnologien, Universit\"{a}t Hamburg, Luruper
Chaussee 149, D-22761 Hamburg, Germany.}
\author{P. Schmelcher}
\affiliation{Zentrum f\"{u}r Optische Quantentechnologien, Universit\"{a}t Hamburg, Luruper
Chaussee 149, D-22761 Hamburg, Germany.}
\author{H. R. Sadeghpour}
\affiliation{ITAMP, Harvard-Smithsonian Center for Astrophysics, Cambridge, MA 02138}

\pacs{33.80.Rv, 31.50.Df}

\begin{abstract}
The internal electric field of a Rydberg atom electron can 
bind a polar molecule to form a giant ultralong-range stable polyatomic
molecule. Such molecules not only share their properties with Rydberg atoms, they possess huge permanent electric dipole moments and in addition allow for coherent control of the polar molecule orientation. In this work, we include additional Rydberg manifolds which couple to the nearly degenerate set of Rydberg states employed in [S. T. Rittenhouse and H. R. Sadeghpour, \prl {\bf104}, 243002 (2010)]. 
The coupling of a set of $(n+3)s$ Rydberg states with the $n(l>2)$ nearly degenerate Rydberg manifolds in alkali metal atoms leads to pronounced avoided crossings in the Born-Oppenheimer potentials. Ultimately,
these avoided crossings enable the formation of the giant polyatomic
Rydberg molecules with standard two-photon laser photoassociation techniques.
\end{abstract}
\date{\today}
\maketitle

\section{Introduction}

The physics of Rydberg atoms and molecules has developed into a
quasi-sustainable ``ecosystem'' over the last decade, mainly due to a) the
exaggerated properties of Rydberg atoms and molecules (long lifetimes, large
sizes, and scalability), and b) exquisite experimental control over these
properties through the advent of ultracold atomic Rydberg samples
\cite{RevModPhys.82.2313,gallagher94}. Some landmark developments over the
last few years in the field of Rydberg physics have been the creation of a
frozen Rydberg gas \cite{PhysRevLett.80.249}, the demonstration of a Rydberg
blockade scheme and the subsequent realization of a Rydberg qubit gate
\cite{Urban2009,Gaetan2009,PhysRevLett.104.010502,PhysRevLett.104.010503},
the formation of ultralong-range isotropic Rydberg molecule \cite{Bendkowsky2009}, 
the creation ultracold neutral plasmas \cite{Killian2007}, as well as the formation of highly-magnetized Rydberg antihydrogen atoms \cite{Pohl2009181}. These rapid advances now allow for
few-body and many-body effects to be realized in laboratory settings and be
prototyped for simulation of strongly interacting spin chains \cite{PhysRevLett.104.043002,PhysRevLett.103.185302,Tezak2010}. 
The other ingredient of our proposed polyatomic Rydberg complex, namely,
ultracold polar molecules, have also been touted as toy models for the simulation
of many-body condensed systems and for the realization of quantum gates
\cite{1367-2630-11-5-055049}. A main appeal of polar molecules is that they may possess sizeable permanent
dipole moments.

In this work, we harness the long-range interaction of a Rydberg atom with a
polar molecule (Rb is used as an ubiquitous example, but other alkali metal atoms would serve the purpose), to demonstrate that ultralong-range polyatomic molecules with enormous permanent dipole
moments can form from combining a Rydberg atom and a polar molecule. 
This system was first studied in \cite{PhysRevLett.104.243002} where it was shown
that the Rydberg electron can be used to ``drag'' the polar molecule into a
preferred orientation and that this process can be controlled by a Raman microwave pulse scheme. 
Ultralong range Rydberg molecules in general came in vogue after a proposal
that such molecules could form from zero-range
interaction of electrons with ground state atoms \cite{PhysRevLett.85.2458}.
The recent experimental realization of a class of such molecules \cite{Bendkowsky2009} has led to revived
theoretical and experimental activity
\cite{PhysRevLett.105.163201,Butscher2010}.

The interaction of the polar perturber with the Rydberg atom strongly couples
the field-free atomic Rydberg states by means of the dipole's electric field.
In \cite{PhysRevLett.104.243002}, the mixing of states was
restricted to the nearly degenerate (negligible quantum defects) manifold of Rydberg states in Rb($nl\gtrsim3$).
In the present work, we extend our previous study by going beyond this
degenerate perturbation theory approach. In this manner, the effect of additional
Rydberg orbitals on the molecular Born-Oppenheimer (BO) potentials is probed.
In particular, the proximity of the $n(l>2)$ Rydberg manifold to a single $(n+3)s$ state introduces a strong coupling, 
allowing for the experimental preparation of the proposed polyatomic Rydberg
molecules via a standard two-photon laser excitation.

The paper is outlined as follows. In section \ref{sec:polarmolec} the model 
Hamiltonian describing the polar molecule perturber is introduced.
The adiabatic Hamiltonian for the full Rydberg atom plus polar molecule complex
is subsequently provided in section \ref{sec:adiabaticH}.
The resulting adiabatic potential surfaces are discussed in section 
\ref{sec:BO}. In section V we discuss the admixture of $s$-wave electron character in the giant Rydberg molecule.
We conclude by a brief summary and an outlook on further research
directions in section \ref{sec:concl}.

\section{Polar molecule model Hamiltonian}\label{sec:polarmolec}

Before we discuss the properties of the proposed polyatomic Rydberg molecules, we must introduce an effective Hamiltonian for the molecular perturber. 
In this context, we model the polar molecule as a
two-level molecule in which the opposite parity states are mixed in the presence of an external field,
i.e.,
\begin{equation}
H_{mol}=\left(
\begin{array}
[c]{cc}%
0 & d_{0}F_{ext}\\
d_{0}F_{ext} & \Delta
\end{array}
\right)  , \label{Eq:H_mol}%
\end{equation}
where $\Delta$ is the zero field splitting between the two molecular states (as in a $\Lambda$-doublet molecule),
$d_{0}$ is the permanent dipole moment of the molecule in the body fixed
frame, and $F_{ext}$ is an electric field external to the polar molecule. Specifically, the external field stems from the Rydberg electron and the Rydberg
core, i.e.,
\begin{equation}
F_{ext}\left(  \vec{R},\vec{r}\right)  =e\left\vert \dfrac{\hat{R}}{R^{2}%
}+\dfrac{\left(  \vec{r}-\vec{R}\right)  }{\left\vert \vec{r}-\vec
{R}\right\vert ^{3}}\right\vert , \label{Eq:Ryd_field}%
\end{equation}
where $e$ is the electron charge, $\vec{R}$ is the core-polar molecule
separation vector and $\vec{r}$ is the position of the Rydberg electron with
respect to the core. In other words, (\ref{Eq:H_mol}) describes the
coupling between the internal state of the polar molecule and the Rydberg atom.

It should be noted that (\ref{Eq:H_mol}) only depends on the magnitude of
the external field, which is appropriate for static, homogeneous fields.
For a rigid rotor molecule, such as
KRb, the rotation of the molecule happens on much slower time scales than
the Rydberg electron orbital time. In this case, the coupling between the Rydberg
atom and the polar molecule accounts for the rotation of the molecule,
\begin{equation}
V_{mol}\left(  \vec{R}\right)  =-\vec{d}\cdot\vec{F}_{ext}, \label{Eq:ed_inter}%
\end{equation}
where $\vec{d}$ is the rigid rotor dipole moment. This interaction
leads to BO potential surfaces which depend on
both $\vec{R}$ and the orientation of $\vec{d}$ with respect to $\vec{R}$;
this more complete treatment is the subject of an ongoing study.
In contrast, a $\Lambda$ doublet molecule, such as OH, has a permanent
dipole moment that arises from the interaction of two opposite parity electronic ($e,f$) states. Rotational transition energies in $\Lambda$-doublet molecules are usually orders of magnitude larger than the typical doublet splitting energies. 
For the
fields provided by highly excited Rydberg atoms, of order $F_{ext}\sim10^{-6}$ a.u., 
our model Hamiltonian (\ref{Eq:H_mol}) thus provides an excellent description of such molecules.

The electron-dipole interaction in (\ref{Eq:ed_inter}) and in the
off-diagonal elements of (\ref{Eq:H_mol}) has a critical value. If the
dipole moment, $d_{0}$, is larger than the Fermi-Teller critical value,
$d_{c}=1.63$ D, an infinite number of bound states form \cite{turner:758}.
Furthermore, an electron scattering off a super critical dipole ($d_{0}>d_{c}$) is
sensitive to the detailed, short-range structure of the polar molecule. To
avoid the complication of super-critical dipole scattering and electron
transfer from the Rydberg atom to the polar molecule, we will deal
exclusively with polar molecules whose dipole moments are subcritical, i.e.,
$d_{0}<1.63$ D. 

\section{The adiabatic Hamiltonian}\label{sec:adiabaticH}

For simplicity, we assume that the external electric field is along the intermolecular axis, such that (\ref{Eq:Ryd_field}) becomes
\begin{equation}
F_{ext}\left(  \vec{R},\vec{r}\right)  =\dfrac{e}{R^{2}}+\dfrac{e\cos
\theta_{\vec{r}-\vec{R}}}{\left\vert \vec{r}-\vec{R}\right\vert ^{2}}
\label{Eq:zaxis_field}%
\end{equation}
where $\theta_{\vec{r}-\vec{R}}=\left(  \vec{r}-\vec{R}\right)  \cdot\vec
{R}/R\left\vert \vec{r}-\vec{R}\right\vert $. This means that the projection $m$
of the Rydberg electron angular momentum along $\vec{R}$ is conserved. To
find the BO potentials, we solve the adiabatic
Schr\"{o}dinger equation at fixed polar molecule location $\vec{R}=R\hat{z}$,
\begin{align}
H_{ad}\psi\left(  R;\vec{r},\sigma\right)   &  =U(R)  \psi(
R;\vec{r},\sigma)  ,\label{Eq:Adiab_SE}
\end{align}
with
\begin{equation}\label{eq:Had}
H_{ad}   = H_A +H_{mol}.
\end{equation}
The first term, $H_A=-\frac{\hbar^{2}}{2m_{e}}\nabla_{r}^{2}+V_l(r)$, describes the unperturbed Rydberg atom; the core penetration, scattering, and polarization effects of its valence electron are accounted for by the $l$-dependent model potential $V_l(r)$ \cite{marinescudca1994}, giving rise to the quantum defects of the low angular momentum Rydberg states. $m_{e}$ is the electron mass, $\psi$ is the electron wave function, and
$\sigma$ is a coordinate signifying the internal states of the polar molecule.
The eigenvalues $U(R)$ serve as the sought-after BO potentials for the polar perturber.
To solve (\ref{Eq:Adiab_SE}), the total wave function is expanded in the basis 
$\{\psi_{nlm}\left(\vec{r}\right) \left\vert \pm\right\rangle\}$ where $\psi_{nlm}\left(  \vec
{r}\right)  $ is an unperturbed Rydberg orbital and $\left\vert \pm
\right\rangle $ are the polar molecule parity states. The atomic and molecular degrees of 
freedom are coupled by the electric field (\ref{Eq:zaxis_field}) that mixes both the Rydberg
orbitals as well as the parity states of the polar molecule, cf.\ (\ref{Eq:H_mol}).

In \cite{PhysRevLett.104.243002}, the BO potentials were found using
degenerate perturbation theory in the (nearly) degenerate set of Rydberg
orbitals $\{\psi_{n\left(  l>2\right)  0}\left(  \vec{r}\right)\}$. 
The molecular Hamiltonian (\ref{Eq:H_mol}) can be prediagonalized to
yield the eigenvalues
\begin{equation}
\varepsilon\left(  F_{ext}\right)  =d_{0}\left(F_{c}\pm\sqrt{F_{ext}%
^{2}+F_{c}^{2}}\right), \label{Eq:Mol_Evals}%
\end{equation}
where $F_{c}=\Delta/2d_{0}$ is the critical external field strength at which
the molecule becomes completely polarized.
Inserting in (\ref{Eq:Mol_Evals}) for $F_{ext}$ the
$R$-dependent eigenstates of the Rydberg atom exposed to the field (\ref{Eq:zaxis_field})
results in the potentials
found in \cite{PhysRevLett.104.243002}.
 A more complete treatment of the
system requires that the total adiabatic Hamiltonian is diagonalized in a
complete set of Rydberg orbitals. Because in the latter case, the orbitals are in general not
degenerate, the prediagonalization scheme cannot be directly used beyond
the degenerate perturbation theory.

In the present work, we go beyond the degenerate perturbation theory
framework adapted in \cite{PhysRevLett.104.243002}. 
To this end, the Hamiltonian (\ref{eq:Had}) is diagonalized in an 
extended basis set comprising several Rydberg $n$ manifolds.
By construction, the atomic Hamiltonian $H_A$ is diagonal in the Rydberg basis $\{\psi_{nl0}(r)\}$,
\begin{align*}
\left\langle \psi_{n^{\prime}l^{\prime}0},\sigma^{\prime}\left\vert -\dfrac
{1}{2}\nabla_{r}^{2}+V_l(r)\right\vert \psi_{nl0},\sigma\right\rangle
&  =\Delta\delta_{\sigma,-}\delta_{n^{\prime}n}\delta_{l^{\prime}l}%
\delta_{\sigma^{\prime}\sigma}\\
&  -\delta_{n^{\prime}n}\delta_{l^{\prime}l}\delta_{\sigma^{\prime}\sigma
}\dfrac{1}{2\left(  n-\mu_{l}\right)  ^{2}},
\end{align*}
where $\mu_{l}$ is the quantum defect for the $l$-th partial wave Rydberg state and
$\sigma$ denotes the parity state of the polar molecule.
The quantum defects used in this work are those for the rubidium atom \cite{gallagher94}:
$\mu_{s}=3.13$, $\mu_{p}=2.65$, $\mu_{d}=1.35$, $\mu_{f}=0.016$, and $\mu_{l>3} \approx0$.
While our focus is on Rb, the method can be extended to any highly
excited Rydberg atom by using the appropriate quantum defects. 

The off-diagonal matrix elements of  (\ref{eq:Had}) are due to $H_{mol}$ which couples the Rydberg electron states to the internal
parity states of the polar molecule. They are determined by evaluating the integrals
\begin{equation}
\left\langle \psi_{n^{\prime}l;0}\left\vert F_{ext}\right\vert \psi
_{nl0}\right\rangle =-e\int d^{3}r\psi_{n^{\prime}l^{\prime}0}^{\ast}\left(
\vec{r}\right)  \dfrac{\cos\theta_{\vec{r}-\vec{R}}}{\left\vert \vec{r}%
-\vec{R}\right\vert ^{2}}\psi_{nl0}\left(  \vec{r}\right).
\label{Eq:ME_Ints}%
\end{equation}
An intuitive closed form of (\ref{Eq:ME_Ints}) is obtained if we expand the electron electric
field contribution in spherical harmonics.
Using the multipole expansion of \cite{Jackson} yields
\begin{equation}
\dfrac{\cos\theta_{\vec{r}-\vec{R}}}{\left\vert \vec{r}-\vec{R}\right\vert
^{2}}=\left\{
\begin{array}
[c]{cc}%
-\sqrt{4\pi}\sum\limits_{l}\dfrac{R^{l-1}}{r^{l+1}}\dfrac{l\sqrt{2l+1}%
}{\left(  2l+1\right)  }Y_{l0}\left(  \theta,\phi\right)  , & r>R\\
\sqrt{4\pi}\sum\limits_{l}\dfrac{r^{l}}{R^{l+2}}\dfrac{\left(  l+1\right)
\sqrt{2l+1}}{\left(  2l+1\right)  }Y_{l0}\left(  \theta,\phi\right)  , & r<R
\end{array}
\right.  .\label{Eq:SH_Expand}%
\end{equation}
We note that, with some generalization, this procedure can also be used to
evaluate the interaction matrix elements between a permanent dipole and the
Rydberg electron, cf.\ (\ref{Eq:ed_inter}). Using (\ref{Eq:SH_Expand}), the
off diagonal matrix elements of (\ref{Eq:Adiab_SE}) are given
by%
\begin{widetext}
\begin{align}
\left\langle \psi_{n^{\prime}l^{\prime}0}\left\vert \dfrac{\cos\theta_{\vec
{r}-\vec{R}}}{\left\vert \vec{r}-\vec{R}\right\vert ^{2}}\right\vert
\psi_{nl0}\right\rangle  &  =\sqrt{\left(  2l^{\prime
}+1\right)  \left(  2l+1\right)  }\sum_{l^{\prime
\prime}=\left\vert l-l^{\prime}\right\vert }^{l+l^{\prime}}\left(
\begin{array}
[c]{ccc}%
l & l^{\prime} & l^{\prime\prime}\\
0 & 0 & 0
\end{array}
\right)^2
\label{Eq:F_MEs}\\
&  \times\left[  \left(  l^{\prime\prime}+1\right)  \dfrac{1}{R^{l^{\prime
\prime}+2}}\int_{0}^{R}r^{l^{\prime\prime}+2}R_{nl}^{\ast}\left(  r\right)
R_{n^{\prime}l^{\prime}}\left(  r\right)  dr-R^{l^{\prime\prime}-1}l^{\prime\prime}%
\int_{R}^{\infty}\dfrac{1}{r^{l^{\prime\prime}-1}}R_{nl}^{\ast}\left(
r\right)  R_{n^{\prime}l^{\prime}}\left(  r\right)  dr\right],  \nonumber
\end{align}
\end{widetext}
where $R_{nl}\left(  r\right)$ is the radial wave function for an
electron in the $nl$ Rydberg state.

The adiabatic Hamiltonian matrix assumes the following form,
\begin{equation}
\bar{H}_{ad}=\left(
\begin{array}
[c]{cc}%
\bar{E}_{Ryd} & d_{0}\bar{F}\\
d_{0}\bar{F} & \Delta\bar{I}+\bar{E}_{Ryd}%
\end{array}
\right),  \label{Eq:Had_mat}%
\end{equation}
where $\bar{I}$ is the unity matrix, $\bar{E}_{Ryd}$ is the diagonal matrix representation of $H_A$,
and $\bar{F}$ is the electric field matrix whose
elements are given by (\ref{Eq:F_MEs}) with a diagonal offset of $1/R^{2}$
due to the contribution of the electric field from the positive ionic core 
to the total electric field, as defined in (\ref{Eq:zaxis_field}). The
BO potentials are found by diagonalizing the matrix (\ref{Eq:Had_mat})
at each $R$. 
Special attention must thereby be drawn to the actual size of the basis set 
in order to ensure convergence of the potentials.

\section{Born-Oppenheimer potentials}\label{sec:BO}

\begin{figure}
\begin{center}
\includegraphics[width=3in]{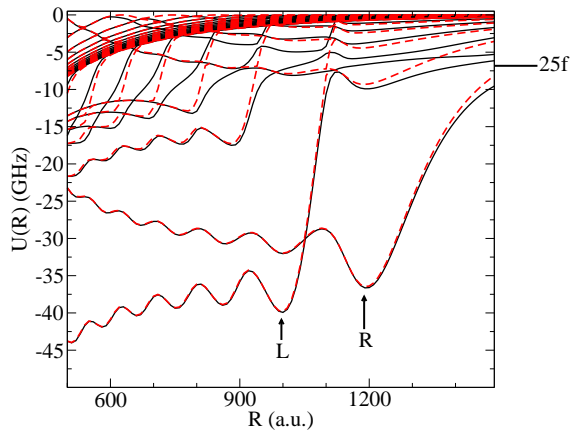}
\end{center}
\caption{The BO potentials for the Rb($n=25$) Rydberg atom and a polar molecule
are shown. The potentials are calculated at the level of degenerate
perturbation theory including (dashed red curves) and ignoring (solid black
curves) the $f$-wave quantum defect of the Rydberg electron. 
The polar perturber in this example is a molecule with a dipole moment
$d_{0}=0.40$ a.u.\ and a zero field splitting $\Delta=1.85\times10^{-7}$ a.u.
 }%
\label{Fig:DPT_vs_f_defect}%
\end{figure}

In \cite{PhysRevLett.104.243002}, it was assumed that for $l>2$, the
Rydberg orbitals of Rb are degenerate and that only this degenerate set of
states is required to converge the BO potentials. Here we extend this treatment
to including additional Rydberg orbitals and the small, but finite, $f$-wave
quantum defect. Figure \ref{Fig:DPT_vs_f_defect} compares the BO
potentials using degenerate perturbation theory (solid curves, as in 
\cite{PhysRevLett.104.243002}) to those found by including the $f$-wave
quantum defect (dotted curves) for the $n=25$ state of rubidium; a polar
molecule dipole moment of $d_{0}=0.40$ a.u., and a zero-field splitting of
$\Delta=1.85\times10^{-7}$ a.u. is considered.

We reiterate some of the most salient features of the proposed molecules.
The modulations that form a series of wells reflect the oscillatory nature of
the Rydberg electron wave function. The outer most wells in the lowest two
potentials are deep enough to support many vibrational levels. The resulting giant polyatomic
Rydberg molecules share several features with previously predicted homonuclear
Rydberg molecules, the so called ``trilobite'' molecules
\cite{PhysRevLett.85.2458,hamilton2002shape,chibisov2002energies}. The size of
these molecules scales as $R_{ryd}\propto n^{2}$ and the well depths scale as
$V_{D}\propto d_{0}/n^{3}$. Unlike for homonuclear Rydberg molecules, the
anisotropic nature of the electron-dipole interaction creates two different
internal configurations. Corresponding to the well labelled R in
figure \ref{Fig:DPT_vs_f_defect}, in one configuration the dipole of the polar molecule is
oriented towards the positive core. In the other configuration -- labelled L -- 
the dipole of the polar molecule is oriented away from the
positive core. 

As anticipated, inclusion of the small $f$-wave quantum defect only modifies the
potentials close to the dissociation threshold limit, i.e., near the Rb($n=25$)
limit. In the lowest wells, where the Rydberg molecules form, the two sets of
potentials are slightly shifted with respect to each other while being
otherwise almost identical, indicating
that including the $f$-wave quantum defect has almost no influence on the
behaviour of the resulting giant molecule, nor on the polyatomic molecular
dipole moment.

By including more Rydberg orbitals in our basis set, we can explore the range of
validity of the degenerate perturbation theory approach.
For smaller dipole
moments, $d_{0}\lesssim0.4$ a.u., relatively few basis states are required
to achieve convergence. In fact, the converged potentials are only slightly
different from those shown in figure \ref{Fig:DPT_vs_f_defect}. As the dipole
moment increases, more and more Rydberg orbitals are required. This is due to
the localizing effects of the dipole-electron potential. For a nearly critical
dipole moment, $d_{0}\approx0.63$ a.u., the electron is almost entirely localized
at the location of the dipole. To accurately describe this (small angle) localization requires an inordinately 
large number of Rydberg orbitals. Furthermore, for larger values of $d_{0}$,
the electron wave function penetrates into the short-range region of the
perturber molecule and probes the detailed structure of the molecule. We therefore restrict here to sub-critical dipole moments,
$d_{0}\leq0.6$ a.u. Because the Rydberg spacing scales as $n^{-3}$, larger sets of orbitals are required
to converge the potentials attached to higher Rydberg thresholds. The calculations reported here are for Rb($n=25$) manifold of states, though the qualitative
behaviours we discuss will persist for higher $n$. To acquire converged
potentials for $n=25$ at the largest dipole moments considered here,
$d_{0}=0.60$ a.u., our basis includes all electron angular momenta for
$n=19,20,...,30,31$ as well as the 32$\left(  s,p,d\right)  $, 33$\left(
s,p\right)  $ and 34s Rydberg orbits, yielding a total of 331 electron basis states. 

\begin{figure}
\begin{center}
\includegraphics[width=3in]{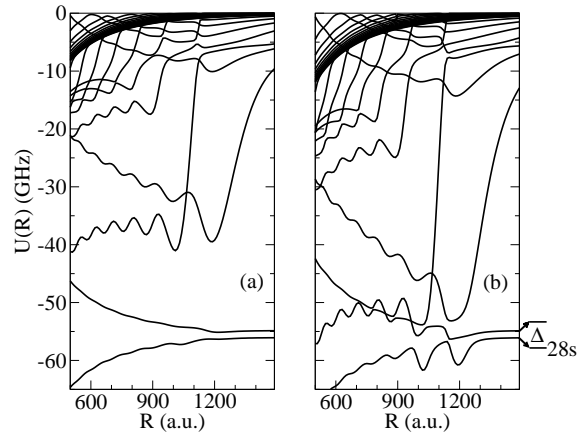}
\end{center}
\caption{
(a) The BO potentials for the Rydberg-polar molecule system are shown,
calculated including lower angular momentum states using a polar molecule
$\Lambda$-doublet parity splitting of $\Delta=1.85\times10^{-7}$ a.u. and a
dipole moment of $d_{0}=0.40$ a.u.\ (subcritical dipole). (b) The same as (a)
with a dipole moment of $d_{0}=0.57$ a.u.\ (the permanent dipole moment of CD).
The zero in energy is the Rb($n=25$) threshold. The 28s level and zero field molecular splitting have also been labelled.}%
\label{Fig:d_0_change_compare}%
\end{figure}

Figure \ref{Fig:d_0_change_compare} shows the converged potentials attached
to the $n=25$ Rydberg level. As in figure \ref{Fig:DPT_vs_f_defect},  the polar
molecule in figure \ref{Fig:d_0_change_compare}(a) has a dipole moment  $d_{0}=0.40$ a.u.\ and a zero field splitting $\Delta=1.85\times10^{-7}$ a.u.\
($\Delta$ has been chosen to be the $\Lambda$-doublet splitting of CD). 
Comparing the two figures shows
that the molecular potentials are unaffected by the presence of
the non-binding $s$-wave states for cases when the molecular dipole is less
than $\sim0.4$ a.u. 
The two curves converging to the Rb($28s$) threshold correspond to the two
opposite parity states of the polar molecule and are correspondingly split by the 
zero field splitting $\Delta$.

For larger dipole moments, it is essential to include the lower angular momentum states in our treatment. While for an unperturbed Rydberg atom, the latter are isolated from the degenerate $l>2$ manifold, increasing the dipole moment has the interesting effect
of forcing the potential wells of the $n=25$ to cross through the $28s$ potentials.
This is possible because the Rb($ns$) quantum defect ($\mu_{s}=3.13$) has 
a small non-integer fraction, placing the Rb($28s$) close to the 
$n=25$ degenerate manifold of states. 
In figure \ref{Fig:d_0_change_compare}(b), a corresponding example 
with $d_{0}$ $=0.57$ a.u.$=1.46$ D -- the permanent dipole moment of CD -- is provided. 
In this case, the interaction between the $n=25$ molecular states and the Rb($28s$) state
is significant. This is generally true of all Rb($n(l>2)$) degenerate manifolds and $(n+3)s$ interacting states. 
As a result and due to strong avoided crossings, potential wells capable of
supporting bound vibrational levels appear in the BO curves attached to the
Rb($28s$) threshold.
Figure \ref{Fig:Diab_pots}(a) provides a close-up of the four 
BO potentials directly involved in the
binding of the polyatomic Rydberg molecule.
The vibrational wave functions for two such bound states in the outermost wells
are also shown.

\section{s-wave admixture}
\begin{figure}[ptb]
\begin{center}
\includegraphics[width=3in]{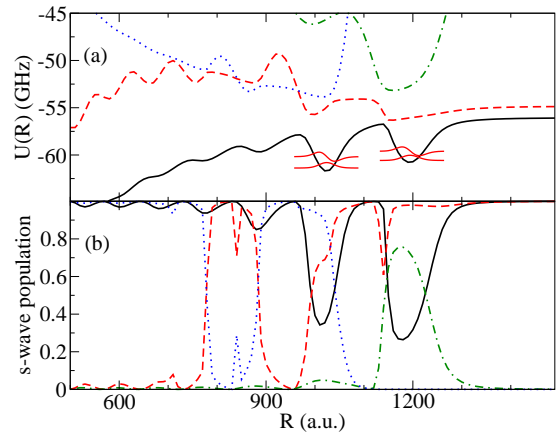}
\end{center}
\caption{(a) The four BO potentials, directly involved in molecular
binding, are shown for the same parameters as in figure \ref{Fig:d_0_change_compare}(b). 
Two vibrational wave functions are also shown at their
binding energies in each of the outermost wells. (b) The $s$-wave character
of the molecular wave function is shown for each of the four 
potentials. 
}%
\label{Fig:Diab_pots}%
\end{figure}

The BO potential curves in figure 1 (also in \cite{PhysRevLett.104.243002}) include a nearly degenerate superposition of atomic orbitals. As such, they solely contain contributions from high angular momentum states ($l>2$) and hence are only accessible in experiments which "photoassociate" these molecules, if and when low-lying angular momenta are admixed into the degenerate manifold. The strong interaction of the $(n+3)s$ Rydberg state with the lowest BO potential curves belonging to the $n(l>2)$ manifolds in figure \ref{Fig:Diab_pots}(b) admixes large amount of $s$-wave character into the wave functions. 

At the potential minimum, the adiabatic channel function is well approximated by
\begin{equation}
\psi\left( r,\chi\right)  =a\left(  d_{0}\right)  \psi_{d}\left(  \vec
{r}\right)  \left\vert \chi_{d}\right\rangle +b\left(  d_{0}\right)  \psi
_{s}\left(  r\right)  \left\vert \chi_{s}\right\rangle
,\label{Eq:Mol_wf_mixture}%
\end{equation}
where $\psi_{d}$ and $\left\vert \chi_{d}\right\rangle $ are the electronic
wave function and internal polar molecule state, respectively, that include
higher electron angular momentum with $l>2$.
$\psi_{s}$ and
$\left\vert \chi_{s}\right\rangle $ are the $s$-wave Rydberg electron wave function
and the corresponding molecular state, respectively. 
The expansion coefficients $a\left(  d_{0}\right) $ and $b\left(  d_{0}\right)$
depend on the dipole moment of the polar perturber as well as on its position $R$.
Figure \ref{Fig:Diab_pots}(b) gives the $s$-wave electron contribution,
$|b(d_0)|^2$, as a function
of the intermolecular separation distance for each of the adiabatic channel
functions corresponding to the potentials shown in figure \ref{Fig:Diab_pots}%
(a). From figure \ref{Fig:Diab_pots}(b) it can be seen that near the minima of
the outer wells, the electronic state of the giant molecule has a significant
$s$-wave character, approximately $30-40\%$. This implies the possibility that
these molecules could be formed in a simple two photon excitation scheme
similar to that used in the experimental realization of homonuclear Rydberg
molecules \cite{Bendkowsky2009}.
The $n=25$ wells shown in figure \ref{Fig:Diab_pots}(a) are deep enough to support bound vibrational
states, the radial wave functions of the first two such states are shown at
their corresponding binding energies in the outer two wells. These molecular levels are well isolated and have readily accessible vibrational frequencies. 

The added $s$-wave character in the electron wave function decreases the charge
localization exhibited in the Rydberg molecule slightly. Even with this
decrease, these Rydberg molecules exhibit massive dipole moments. Using the
scaling behaviour of \cite{PhysRevLett.104.243002}, the Rydberg molecule
dipole moment $d_{Ryd}$ can be found,%
\[
d_{Ryd}\approx1.3\left[  a\left(  d_{0}\right)  \right]  ^{2}n^{2}.
\]
For the case shown in figure \ref{Fig:Diab_pots}(a) this yields $d_{Ryd}%
\approx\allowbreak1400$D, a truly large dipole moment. The presence of these
enormous dipole moments indicates that the polyatomic Rydberg molecules could be sensitively
controlled through the use of small external electric fields. Due to the
extreme sensitivity of Rydberg electron to external fields, the behaviour of
the BO potentials under the influence of such fields is not immediately
obvious and is the subject of ongoing studies.

\begin{figure}
\begin{center}
\includegraphics[width=3.2in]{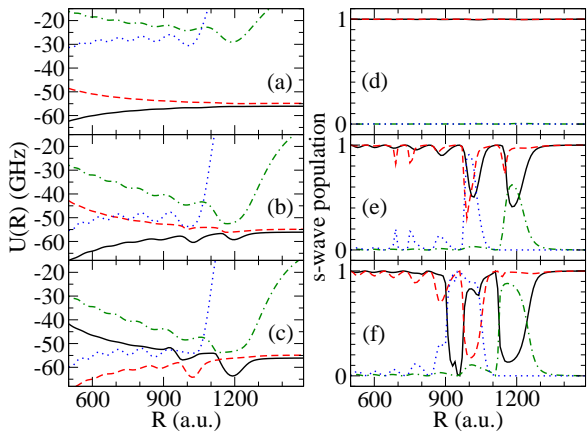}
\end{center}
\caption{The variations of the BO potential curves with the strength of the
polar molecule dipole moment $d_{0}$. (a) $d_{0}=0.30$ a.u., (b)
$d_{0}=0.55$ a.u., and (c) $d_{0}=0.60$ a.u. The respective $s$-wave
contributions to the molecular wave functions are shown in panels (d-f). 
A $\Lambda$-doublet parity splitting of $\Delta=1.85\times10^{-7}$ a.u.\ is considered.
}%
\label{Fig:d0_changes}%
\end{figure}

In figure \ref{Fig:d0_changes}(a-c), we provide snapshots of additional
BO potentials responsible for molecular binding, namely, for
$d_{0}=0.30$, $0.55$, and $0.60$ a.u., respectively.
As the dipole moment of the polar perturber
increases, the L and R wells are pulled down through the $s$-wave Rydberg
threshold. At $d_{0}=0.30$ a.u., the well structure is not perturbed at all by
the $s$-wave potential. For $d_{0}=0.55$, the potential wells are strongly
distorted by the presence of the lower threshold, while for $d_{0}=0.60$ a.u.\
the outermost well structures have passed through the $s$-wave threshold and correspondingly
have a much smaller $s$-wave contribution. Figures \ref{Fig:d0_changes}(d-f)
show the $s$-wave electron contribution for each potential. 
For $0.5$ a.u.$\lesssim d_{0}\lesssim 0.6$ a.u. the $s$-wave character at
the position of the outermost wells varies approximately $10-40\%$, enough to give a fairly large
Rabi frequency for the creation of Rydberg molecules in a standard two photon excitation scheme.

\section{Conclusions and outlook}\label{sec:concl}

In this paper we have examined the formation of giant, polyatomic Rydberg
molecules consisting of a $\Lambda$-doublet polar molecule and a Rydberg atom.
We provided a more complete description of the polar molecule model initially used to predict these nearly macroscopic molecules \cite{PhysRevLett.104.243002}.
By using a two-state model that is sensitive to
external fields, a Rydberg electron is coupled to the internal state of the
polar molecule. The resulting electron-molecule interaction creates a series
of BO potentials with an oscillating series of wells that
reflect the Rydberg oscillations in the electron wave function. By extending
the work in \cite{PhysRevLett.104.243002} to beyond the degenerate
perturbation theory, we have shown that the small, but finite $f$-wave
quantum defect changes little the behaviour of the resulting giant molecules.

By including lower electron-angular momentum states, a new set of potentials
attached to the $\left(  n+3\right)  s$ Rydberg level appear. These added
potentials have no significant influence on electronic state of the Rydberg molecule
for smaller subcritical dipole moments, $d_{0}\lesssim0.4$ a.u. For nearly critical dipole
moments, $d_{0}\approx0.6$ a.u., a series of avoided and level crossings is
formed between the potentials formed from the degenerate $n(l>2)$ manifold and the $(n+3)s$-wave
potentials. These avoided crossings lend significant $s$-wave character to the
Rydberg electron state opening the possibility of creating the giant polyatomic
Rydberg molecules using standard two-photon Rydberg excitation schemes.
The series of avoided and level crossings
create a complex structure of couplings between the various BO potentials.
Extensions of this work beyond a simple
two-state polar molecule to incorporate rigid rotor-type polar molecules, such
as KRb, poses a significant challenge. The added angular behaviour of the polar
molecule will create a set of two-dimensional potential surfaces which couple
the rotational behaviour of the polar molecule to the vibrational state of the
giant Rydberg molecule. Creating and examining this intricate energy landscape
will be the subject of future work.

\section{Acknowledgements}

The authors would like to thank T.~V.~Tscherbul for help in calculating the $\Lambda$-doublet molecule parameters used in this work. MM acknowledges financial support from the German Academic Exchange Service (DAAD). 
STR and HRS acknowledge financial support from the NSF through ITAMP 
at Harvard University and the Smithsonian Astrophysical Observatory.


\end{document}